\documentclass[aps,prstab,reprint,groupedaddress]{revtex4-1}

\usepackage{graphicx}
\usepackage[fleqn]{amsmath} 

\newcommand{\sn}{\mathrm{sn}}
\newcommand{\cn}{\mathrm{cn}}

\newcommand{\sgn}{\mathrm{sgn}}

\begin{document}

\title{Analytical Model of an Isolated Single-atom Electron Source}

\author{W.J. Engelen}
\author{E.J.D. Vredenbregt}
\author{O.J. Luiten}
\email[]{o.j.luiten@tue.nl}
\affiliation{Department of Applied Physics, Eindhoven University of Technology, P.O. Box 513, 5600 MB Eindhoven, The Netherlands}

\date{\today}

\begin{abstract}
An analytical model of a single-atom electron source is presented, where electrons are created by near-threshold photoionization of an isolated atom. The model considers the classical dynamics of the electron just after the photon absorption, i.e. its motion in the potential of a singly charged ion and a uniform electric field used for acceleration. From closed expressions for the asymptotic transverse electron velocities and trajectories, the effective source temperature and the virtual source size can be calculated. The influence of the acceleration field strength and the ionization laser energy on these properties has been studied. With this model, a single-atom electron source with the optimum electron beam properties can be designed. Furthermore, we show that the model is also applicable to ionization of rubidium atoms, and thus also describes the ultracold electron source, which is based on photoionization of laser-cooled alkali atoms. 
\end{abstract}

\maketitle

\section{Introduction}
Single-atom electron emitters are the closest one can get to the ideal of a point-like electron source. They are capable of producing beams with nearly full spatial coherence, i.e. approaching the Heisenberg uncertainty limit $\sigma_x \sigma_{p_x} = \hbar / 2$, with $\sigma_x$ the root-mean-square (rms) transverse beam size and $\sigma_{p_x}$ the rms transverse momentum spread; this was recently demonstrated experimentally \cite{Mutus_NJP_13, Chang_N_09}. Currents in the range of tens of nA can be generated in this way, resulting in ultra-bright electron beams. The fabrication of single-atom emitters, however, involves complicated preparation methods, usually based on the formation of a tiny protrusion on the apex of a 10--100\,nm hemispherical metallic tip \cite{Fink_IJRD_86, Kuo_NL_04}. After proper treatment the protrusion forms a sharp structure with a single atom on top, which sprays a narrow beam of electrons when an external field is applied. New preparation methods have yielded relatively robust devices \cite{Kuo_NL_04,Rezeq_JCP_06}, but issues of stability, reproducibility and lifetime remain.

In the past few years, a new type of electron source has emerged, based on near-threshold photoionization of a laser-cooled and trapped atomic gas \cite{Engelen_NC_13, Engelen_U_14, Engelen_NJP_13, McCulloch_NC_13, McCulloch_NP_11, Saliba_OE_12, Taban_EPL_10}, which may provide an interesting alternative. In this ultracold source, the atoms are suspended in vacuum, virtually motionless, and only emit an electron when triggered by light of the proper wavelength. They thus constitute an entire new class of single-atom emitters with unique properties: all emitters are exactly identical and reproducible; emission is on demand, allowing ultrafast operation; after emission the remaining ion is discarded, i.e., each shot a fresh source is used and thus no ageing problems will occur; many atoms may be ionized in parallel, the equivalent of an array of conventional single-atom emitters; as will be discussed in more detail in this work, the combination of applied electric field strength and ionization wavelength enables precise tailoring of the emission angular spread and the virtual source size.

Here we discuss a model of a single-alkali-atom emitter, freely suspended in vacuum, in terms of a theoretical description of the ionization process. The basis is the exactly solvable classical, non-relativistic model for ionization of the hydrogen atom \cite{Kondratovich_JPB_84a, Kondratovich_JPB_84b, Bordas_PRA_98}. We use a classical model, as the motion of a quasifree electron is reasonably well described by classical trajectories \cite{Bordas_PRA_98} and we are not interested in interference effects, for which a full quantum wave-packet description is needed. 
Recently, this model has been used to simulate images recorded with photoelectron imaging spectroscopy, a technique for studying the velocity and angular distribution of electrons created by photoionization \cite{Stodolna_PRL_13, Bordas_PRA_98, Nicole_PRL_00, Bordas_PRA_03, Lepine_PRA_04}. 
We expand on this description by deriving closed analytical expressions for the asymptotic transverse momentum distribution of the emitted electron beam and the virtual source size, in terms of the applied electric field strength and the ionization wavelength. The rms transverse momentum spread of the electrons $\sigma_{p_x}$ determines the quality of the source; for convenience this will be expressed in this work in terms of the effective transverse source temperature
\begin{equation}
	T \equiv m \sigma_{v_x}^2 / k_{\rm{B}},\label{eq:T_sigma_vx} 
\end{equation}
with $m$ the electron rest mass, $\sigma_{v_x}$ the velocity spread of the electron bunch in the $x$-direction, and $k_{\rm{B}}$ the Boltzmann constant, although typically the velocities in the electron beam do not follow a Maxwell--Boltzmann distribution. We have used this model to explain experimental data of the effective source temperature of an ultracold electron source, based on photoionization of laser-cooled rubidium atoms \cite{Engelen_NC_13, Engelen_U_14, Engelen_NJP_13}. Furthermore, it can be used to design a single-atom electron source with the optimum electron beam properties.

The paper is organized as follows. In Sec. \ref{sec:mod_analyt_model}, the equations describing the electron's classical motion in a Coulomb-Stark potential are treated. These will be used in Sec. \ref{sec:mod_trajectories} to calculate electron trajectories. In Sec. \ref{sec:mod_vr}, closed expressions are derived for the asymptotic transverse electron velocity and the asymptotic parabolic trajectory for a given initial emission angle. By proper averaging over this angle, the transverse velocity spread of the electron beam, and thus the effective source temperature, and the virtual source size are calculated (Sec. \ref{sec:mod_src_temp}). The analytical expressions describing the asymptotic motion of the electrons depend parametrically on the initial emission angle. In Sec. \ref{sec:phase_space} we show how this allows plotting of the $x$-$p_x$ transverse phase space distribution of the beam for a given acceleration field strength and ionization laser energy. On the basis of phase space plots, we discuss the conditions required for realizing a fully coherent beam, only limited by the Heisenberg uncertainty relation, which constitutes the applicability limit of this classical model. The analytical description of the electron velocities and trajectories strictly holds only for hydrogen atoms. However, in Sec. \ref{sec:mod_rubidium_model} we show by particle tracking simulations that the beam produced by near-threshold photoionization of rubidium atoms is accurately described by the hydrogen model as well. We expect this to hold for all alkali species, because of their hydrogen-like electronic structure. This is important in view of the fact that the ultracold source is based on laser-cooled alkali atoms. Finally, in Sec. \ref{sec:mod_ultrashort_model}, we briefly discuss what the source characteristics are when ultrashort, broadband laser pulses are used for photoionization. 
We will end with conclusions and suggestions for further investigations (Sec. \ref{sec:mod_concl}).

\section{Analytical model\label{sec:mod_analyt_model}}
In this work, the classical dynamics of the electron just after the photon absorption in the ionization process is considered. In this process, the electrons gain excess energy $E_{\rm{exc}}$, given by
\begin{equation}
	E_{\rm{exc}} = E_\lambda + E_F,
\end{equation}
with
\begin{equation}
	E_\lambda = h c \left(\frac{1}{\lambda} - \frac{1}{\lambda_0} \right),
\end{equation}
the ionization energy with respect to the zero-field ionization threshold, and 
\begin{equation}
	E_F = 2 H \sqrt{\frac{|F_{\rm{acc}}|}{F_0}},
\end{equation}
the Stark shift of the ionization threshold caused by the applied electric field $F_{\rm{acc}}$ in the accelerator. Here, $\lambda$ is the ionization laser wavelength, $\lambda_0$ is the zero-field ionization wavelength, $H = 27.2$\,eV the Hartree energy, and $F_0 = 5.14 \cdot 10^{11}$\,V/m the atomic unit of field strength. Unless stated otherwise, a field strength $F_{\rm{acc}} = -0.155$\,MV/m and an excess energy $E_{\rm{exc}}$ = 5\,meV are used in this work (so $E_F = 29.9$\,meV and $E_\lambda = -24.9$\,meV) for numerical examples and illustrative plots. These values correspond to typical  parameters of experiments of near-threshold photoionization of laser-cooled atoms \cite{Engelen_NC_13, Engelen_U_14, Engelen_NJP_13, Taban_EPL_10}. 

After the ionization process, the electron will move in the potential of a singly charged ion and a uniform electric field. The corresponding potential energy $U$ is given by 
\begin{equation}
	U = U_{\rm{ion}} + F z,\label{eq:pot_energy}
\end{equation}
with $U_{\rm{ion}}$ the potential energy in the field of the ion and $F = F_{\rm{acc}}/F_0$ the electric field strength. We assume that $F$ is pointing in the $z$-direction, so for negative field strengths electrons escape in the positive $z$-direction. In Eq. \eqref{eq:pot_energy}, and in the remainder of this paper, atomic units are used, unless stated otherwise.

\begin{figure}
	\centering
	\includegraphics{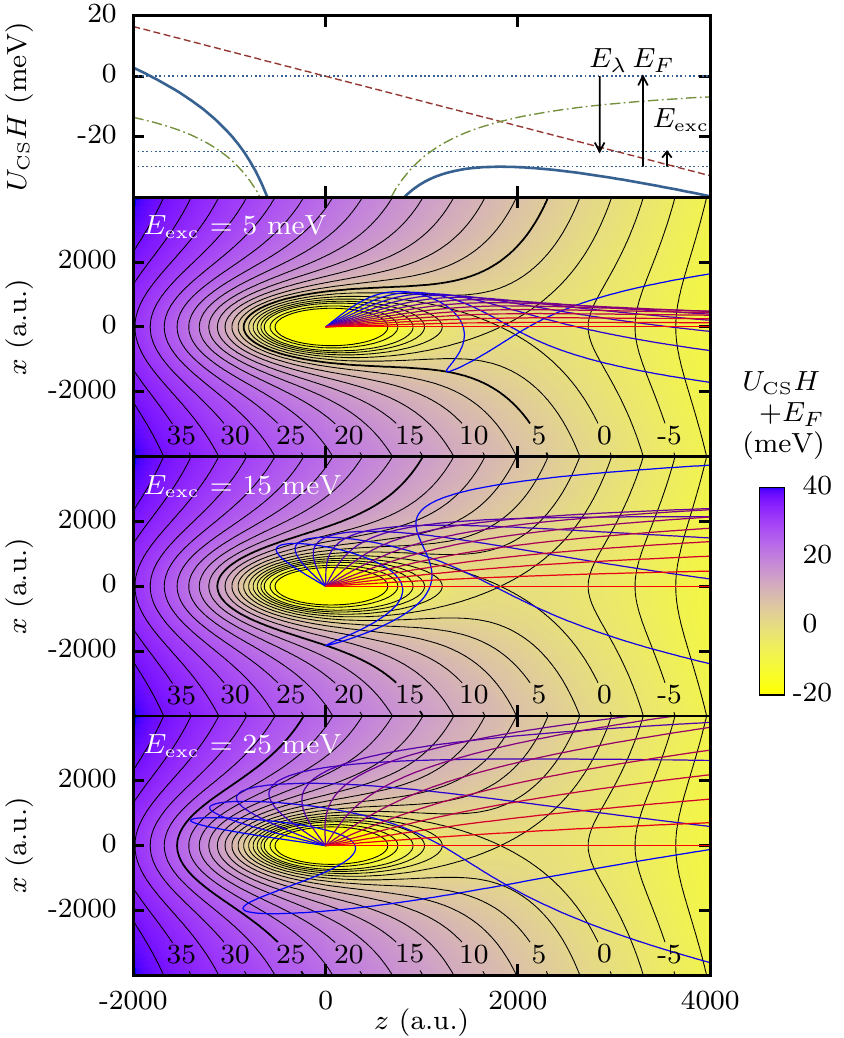}%
	\caption{Top: Combined Coulomb-Stark potential $U_{\rm{CS}}$ (solid curve), together with the Coulomb (dashed-dotted curve) and Stark potential (dashed curve) for $E_{\rm{exc}}$ = 5\,meV and $F_{\rm{acc}} = -0.155$\,MV/m. Bottom: Corresponding potential landscape with equipotential lines, together with electron trajectories for $E_{\rm{exc}} =$ 5, 15, and 25\,meV and different starting angles $0 \le \beta < \beta_{\rm{c}}$, where the colour of the trajectory indicates $\beta$.\label{fig:potential_surf}}
\end{figure}

Using classical mechanics, an analytical solution of the electron's motion in a Coulomb-Stark potential $U_{\rm{CS}}$ can be derived  \cite{Kondratovich_JPB_84a, Kondratovich_JPB_84b, Bordas_PRA_98}. This potential is given by Eq. \ref{eq:pot_energy} with $U_{\rm{ion}} = -1/R$, where $R = \sqrt{x^2 + y^2 + z^2}$ the distance to the ion core. In Figure \ref{fig:potential_surf}, $U_{\rm{CS}}$ is sketched (solid curve) for the standard simulation conditions $E_{\rm{exc}} = $ 5\,meV and $F_{\rm{acc}} = -0.155$\,MV/m. 

In the remainder of this Section, we recall the solution of the electron dynamics in $U_{\rm{CS}}$ \cite{Kondratovich_JPB_84a, Kondratovich_JPB_84b, Bordas_PRA_98}, where the electron has a kinetic energy $E_{\rm{k}} = E - U_{\textrm{CS}}$, with $E = E_\lambda/H$ the zero-field ionization energy. This solution is expressed in parabolic coordinates $\xi$ and $\eta$, with $r = \sqrt{x^2 + y^2} = \sqrt{\xi \eta}$ and $z = (\xi - \eta)/2$.

The motion along the $\xi$-coordinate is given by
\begin{equation}
	\xi(\tau) = \frac{\xi_{\_} \sn^2(\varphi \tau | m_{\xi})}{m_{\xi}^{-1} - \sn^2(\varphi \tau | m_{\xi})},
\end{equation}
with sn a Jacobi elliptic function \cite{Abramowitz_Book_65} and $\tau$ the reduced time variable. This reduced time is related to the real time $t$ by ${\rm{d}}\tau = {\rm{d}}t/R$. At $\tau = 0$, the electron is in the force center. Furthermore, we define
\begin{align}
	&\xi_{\_} = \frac{|E|}{F} \left[ \sqrt{1+\frac{Z_1}{Z_c}} - \sgn(E) \right], \nonumber \\
	&\varphi = \sqrt{\frac{|E|}{2}} \left( 1 + \frac{Z_1}{Z_c} \right)^{1/4}, \nonumber \\
	&m_{\xi} = \frac{1}{2} \left[ 1 + \sgn(E) \left( 1 + \frac{Z_1}{Z_c} \right)^{-1/2} \right], \nonumber \\
	&Z_1 = \sin^2(\beta/2),
	\quad Z_c = \frac{E^2}{4 |F|}.
	\label{eq:xi_defs}
\end{align}
Here $\beta$ is the angle between the initial electron velocity and the $z$-axis. Along the $\xi$-coordinate, the motion is periodic in $\tau$ and thus bound in this direction.

For the motion along the $\eta$-coordinate, three energy regimes can be identified. For $E \leq -E_F/H$ (i.e. $E_{\rm{exc}} \leq 0$), the motion is bound in $\eta$ and no ionization occurs. For $-E_F/H < E \leq 0$, the motion is open if $\beta$ is  smaller than the critical angle $\beta_{\rm{c}}$, given by
\begin{equation}
	\beta_{\rm{c}} = 2 \arccos \left( \frac{-E_{\lambda}}{E_F} \right) = 2 \arccos \left( \frac{-E}{2 \sqrt{|F|}} \right),
	\label{eq:critical_angle}
\end{equation}
and closed otherwise. For $E > 0$, the motion is always open (so $\beta_{\rm{c}} = \pi$), as the electron's energy is above the field-free ionization threshold. For open motion, $\eta$ increases monotonically in time. For $|E| \leq 2 \sqrt{Z_2 |F|}$
\begin{equation}
	\eta(\tau) = m_+ \left( \frac{1- \cn(\theta \tau | m_\eta)}{\sn(\theta \tau | m_\eta)} \right)^2,
	\label{eq:eta_I}
\end{equation}
with cn a Jacobi elliptic function \cite{Abramowitz_Book_65} and
\begin{align}
	&m_+ = \frac{|E|}{F} \sqrt{\frac{Z_2}{Z_c}}, \quad \theta = \sqrt{|E|} \left(\frac{Z_2}{Z_c} \right)^{1/4}, \nonumber \\
	&m_{\eta} = \frac{1}{2} \left[ 1 - \sgn(E) \left( \frac{Z_2}{Z_c} \right)^{-1/2} \right], \nonumber \\
	&Z_2 = \cos^2(\beta/2).
	\label{eq:eta1_defs}
\end{align}
For $E > 2 \sqrt{Z_2 |F|}$
\begin{equation}
	\eta(\tau) = n_+ \frac{\sn^2(\psi \tau | n_\eta)}{1 - \sn^2(\psi \tau | n_\eta)},
	\label{eq:eta2}
\end{equation}
with
\begin{align}
	&n_+ = \frac{E}{F} \left[ 1 - \sqrt{1 - \frac{Z_2}{Z_c}} \right], \nonumber \\
	&\psi = \frac{\sqrt{E}}{2} \sqrt{1 + \sqrt{1 - \frac{Z_2}{Z_c}}}, \nonumber \\
	&n_\eta = 2 \left[ 1 + \left(1 - \frac{Z_2}{Z_c} \right)^{-1/2} \right]^{-1}.
	\label{eq:eta2_defs}
\end{align}

\section{Electron trajectories\label{sec:mod_trajectories}}
With these equations, trajectories can be calculated for electrons leaving the atom. This is illustrated in Fig. \ref{fig:potential_surf}, where trajectories are shown for different starting angles $\beta$ for $E_{\rm{exc}}$ = 5, 15, and 25\,meV, together with the corresponding potential landscape. The trajectories are transformed from the $r$--$z$-plane to the $x$--$z$-plane, so the electron trajectory is continuous in case of a $z$-axis crossing. The electron trajectory corresponding to $\beta$ just below $\beta_{\rm{c}}$ touches the equipotential line $U_{\rm{CS}} H + E_F = E_{\rm{exc}}$ (in SI units); electrons with $\beta \ge \beta_{\rm{c}}$ do not have enough energy to cross that equipotential line and are bound. From the trajectories it can be seen that electrons that are leaving the atom experience a force towards the $z$-axis because of the shape of the potential, which acts as a bottleneck near the saddle-point at $z = -\sgn(F) / \sqrt{|F|}$. This reduces the transverse velocity of the electrons; as $\sigma_{v_r}$ is also reduced, this leads to a lower electron temperature compared to the simple model based on equipartition, given by $T = 2/(3 k) E_{\rm{exc}}$ (in SI units). The simple model would apply for electrons following ballistic (parabolic) trajectories, which happens when the Coulomb interaction is left out (see Appendix \ref{sec:parabol_traj}). For low excess energies, electrons are leaving the atom from a small cone, as seen far from the atom. By increasing the excess energy, the size of the bottleneck is increased; this leads to a larger cone of escaping electrons and thus a higher electron source temperature.

\section{Asymptotic electron velocities and trajectories\label{sec:mod_vr}}

From the expressions for $\xi$ and $\eta$, the transverse velocity of an electron can be calculated
\begin{equation}
	\frac{{\rm{d}}r}{{\rm{d}}t} = \frac{\frac{{\rm{d}}\eta}{{\rm{d}}\tau} \xi + \eta \frac{{\rm{d}}\xi}{{\rm{d}}\tau}}{\sqrt{\eta \xi} \left( \eta + \xi \right)},
\end{equation}
where we used ${\rm{d}}t = {\rm{d}}\tau (\eta + \xi)/2$. With this equation, we are now going to derive the asymptotic transverse velocity $v_r(E, F, \beta)$ of the electrons, which can be used to calculate the temperature.

For $|E| \leq 2 \sqrt{Z_2 |F|}$, the electron escapes to infinity as $\eta \to -\infty$, which is the case when in Eq. \eqref{eq:eta_I} $\sn(\theta \tau | m_\eta) \to 0$. As $\sn\left(2 K(m) | m\right) = 0$, with $K(m)$ the complete elliptic integral of the first kind, and $\cn\left(2 K(m) | m\right) = -1$ \cite{Abramowitz_Book_65}, we can calculate $v_r$ by taking the limit of $\tau \to \tau_{{\rm{max}},m} = 2 K(m_\eta) / \theta$ for ${\rm{d}}r/{\rm{d}}t$. This evaluates to
\begin{equation}
	v_r = 
	\theta \sqrt{ \frac{1}{m_+} \frac{\xi_{\_} m_\xi m_{\mathrm{sn}}^2 }{(1-m_\xi m_{\mathrm{sn}}^2 )} },
	\label{eq:vr_theta}
\end{equation}
with
\begin{equation}
	m_{\mathrm{sn}} = \sn \left(2 \varphi K(m_\eta)/\theta | m_\xi \right).
	\label{eq:theta_sn}
\end{equation}
Using Eqs. \eqref{eq:xi_defs} and \eqref{eq:eta1_defs}, Eq. \eqref{eq:vr_theta} can be written as
\begin{equation}
	v_r = \sqrt{\frac{ 2 |F| m_{\mathrm{sn}}^2 (\cos \beta - 1)}
	{E m_{\mathrm{sn}}^2 + (m_{\mathrm{sn}}^2 - 2) \sqrt{E^2 - 2 |F| (\cos \beta - 1)} } }.
	\label{eq:vr1_simple}
\end{equation}
For the special case $E=0$, $v_r$ can be calculated with Eq. \eqref{eq:vr1_simple} and
\begin{equation}
	m_{\mathrm{sn},E=0} = \sn \left(
	\frac{\sqrt{\pi} (1 - \cos \beta)^{\frac{1}{4}} \Gamma(\frac{1}{4}) }
	{\phantom{\sqrt{}}2 (1 + \cos \beta)^{\frac{1}{4}} \Gamma(\frac{3}{4})} \Big|
	\frac{1}{2} \right),
\end{equation}
with $\Gamma$ the gamma function.

For $E > 2 \sqrt{Z_2 |F|}$, the asymptotic value $v_r$ is reached when $\sn(\psi \tau | n_\eta) \to 1$ in Eq. \eqref{eq:eta2}. As $\sn\left(K(m) | m\right) = 1$ \cite{Abramowitz_Book_65}, we can calculate $v_r$ by taking the limit of $\tau \to \tau_{{\rm{max}},n} = K(n_\eta) / \psi$ for ${\rm{d}}r/{\rm{d}}t$. This evaluates to
\begin{equation}
	v_r = 2 \psi 
	\sqrt{\frac{(1 - n_\eta)}{n_+} \frac{\xi_{\_} m_\xi n_{\mathrm{sn}}^2 }{(1 - m_\xi n_{\mathrm{sn}}^2)}},
	\label{eq:vr_psi}
\end{equation}
with
\begin{equation}
	n_{\mathrm{sn}} = \sn( \varphi K(n_\eta)/\psi | m_\xi).
	\label{eq:psi_sn}
\end{equation}
After substituting Eqs. \eqref{eq:xi_defs} and \eqref{eq:eta2_defs} in Eq. \eqref{eq:vr_psi}, we obtain
\begin{equation}
	v_r = \sqrt{\frac{ 2 |F| n_{\mathrm{sn}}^2 (\cos \beta - 1)}
	{E n_{\mathrm{sn}}^2 + (n_{\mathrm{sn}}^2 - 2) \sqrt{E^2 - 2 |F| (\cos \beta - 1)} } },
\end{equation}
which is very similar to Eq. \eqref{eq:vr1_simple}: only $m_{\mathrm{sn}}$ has been replaced by $n_{\mathrm{sn}}$.

\begin{figure}[tp]
	\centering
	\includegraphics{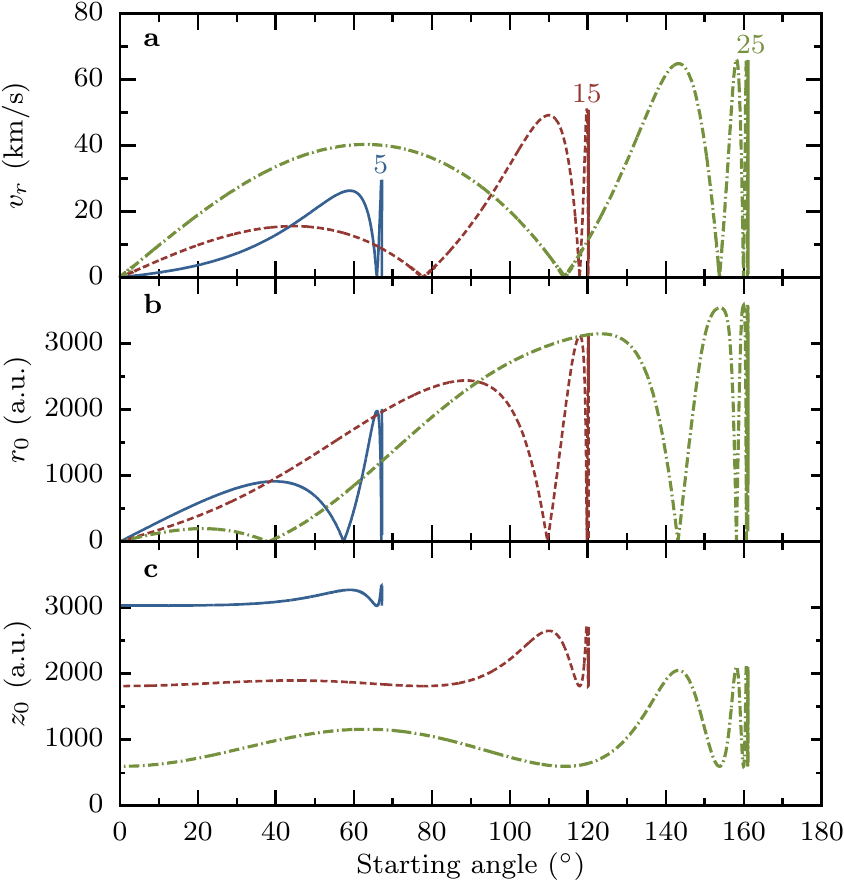}%
	\caption{Asymptotic transverse electron velocity $v_r$ (a), starting position $r_0$ (b), and starting position $z_0$ (c) as a function of the electron's starting angle $\beta$ for $F_{\rm{acc}} = -0.155$\,MV/m and $E_{\rm{exc}} = 5$\,meV (blue solid curve), $E_{\rm{exc}} = 15$\,meV (red dashed curve), and $E_{\rm{exc}} = 25$\,meV (green dashed-dotted curve).\label{fig:vr_r0_z0_vs_beta}}
\end{figure}

\begin{figure}[tp]
	\includegraphics{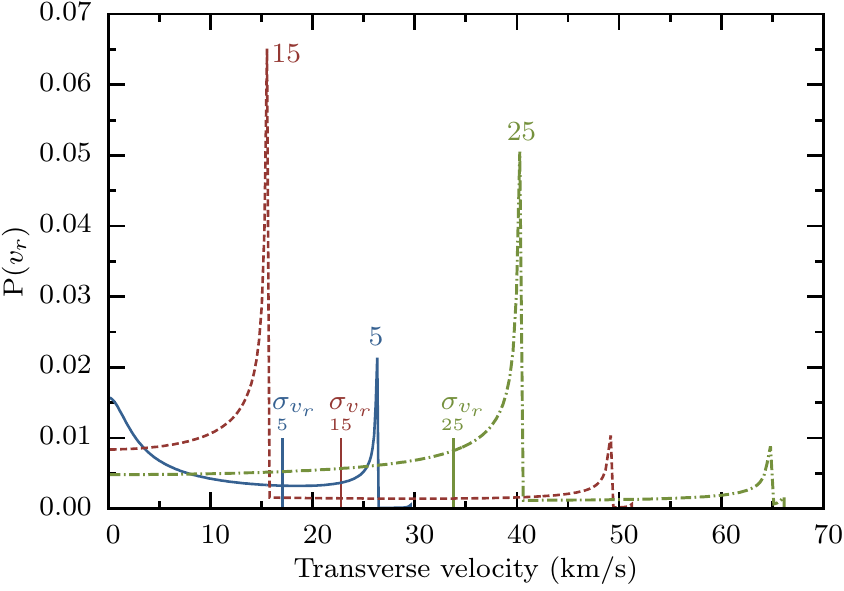}%
	\caption{Asymptotic transverse velocity distribution P$(v_r)$ for $F_{\rm{acc}} = -0.155$\,MV/m and $E_{\rm{exc}} = 5$\,meV (blue solid curve), $E_{\rm{exc}} = 15$\,meV (red dashed curve), and $E_{\rm{exc}} = 25$\,meV (green dashed-dotted curve). Also shown are the transverse velocity spread $\sigma_{v_{r}}$ for these energies (vertical lines).\label{fig:vr_histogram}}
\end{figure}

With these equations, the electron's asymptotic transverse velocity has been calculated as a function of the starting angle $\beta$ for $E_{\rm{exc}}$ = 5, 15, and 25\,meV (Fig. \ref{fig:vr_r0_z0_vs_beta}a). Recall that no electrons can escape above the critical angle $\beta_{\rm{c}}$ and that this angle increases with excess energy. For small $\beta$, $v_r$ increases with $\beta$, as expected. For larger $\beta$, the particles are launched more and more uphill in the potential; as a result, a part of the electron's transverse velocity is converted into longitudinal velocity, leading to a decrease of $v_r$ with increasing $\beta$. This continues until $v_r = 0$, which happens when $\tau = n T_\xi$, with $n$ an integer and $T_{\xi} = 2 \sqrt{2} K(m_{\xi}) \left( 1 + \frac{Z_1}{Z_c} \right)^{-1/4} /\sqrt{|E|}$ the oscillation period of $\xi$ \cite{Bordas_PRA_03}. As the particle escapes to infinity for a finite value of $\tau$, $n$ is also finite. When increasing $\beta$ further,  $v_r$ will go up again, and the corresponding electron trajectory will cross the $z$-axis once. Depending on the value of the excess energy, this process repeats itself once or multiple times, as can be clearly seen in the graph for $E_{\rm{exc}} =$ 15 and 25\,meV. Close to the critical angle, the electrons follow a complex trajectory through the potential, leading to a strong dependence of $v_r$ on the angle. Assuming isotropic emission, the probability distribution P($v_r$) of the transverse velocity has been calculated (see Fig. \ref{fig:vr_histogram}). The local maxima in Fig. \ref{fig:vr_r0_z0_vs_beta}a lead to peaks in the velocity distribution.

\begin{figure}[tp]
	\includegraphics{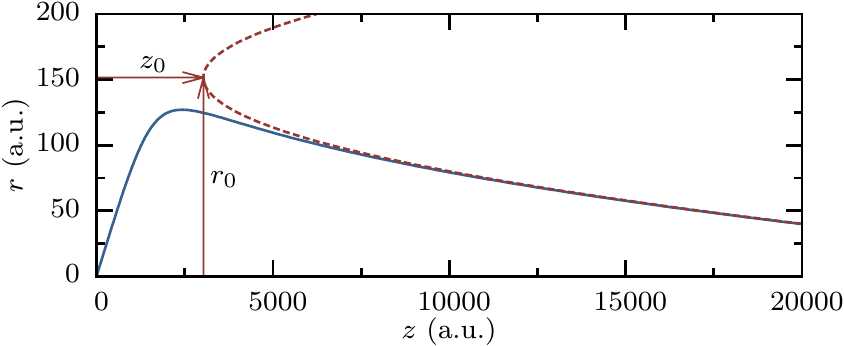}%
	\caption{Electron trajectory (blue solid curve) and the parabola that describes the trajectory in the limit $\tau \to \tau_{\rm{max}}$ (red dashed curve) for $\beta = 5^{\circ}$, $E_{\rm{exc}} =$ 5\,meV, and $F_{\rm{acc}} = -0.155$\,MV/m.\label{fig:eff_src}}
\end{figure}

In the limit of $\tau \to \tau_{\rm{max}}$, the effect of the Coulomb field on the electron's motion can be neglected. The electron trajectory can then be described by the parabola
\begin{equation}
	z - z_0 = a (r - r_0)^2,\label{eq:par_traj}
\end{equation}
with $z_0$ and $r_0$ the apparent starting position of the electron and $a = -F / (2 v_r^2)$. Or to put it in other words: far from the ion core, the asymptotic value $v_r$ is reached, and the electric field only influences $v_z$ and the slope of the trajectory. An example of an electron trajectory and the corresponding parabola for the standard simulation conditions and $\beta = 5^{\circ}$ is shown in Fig. \ref{fig:eff_src}.
In the remainder of this Section, we derive expressions for $r_0$ and $z_0$, which allow us to calculate the virtual source size. 
We first rewrite Eq. \eqref{eq:par_traj} as	$r_0 = r - \frac{1}{2 a} \frac{{\rm{d}}z}{{\rm{d}}r}$.
In the limit $\tau \to \tau_{\rm{max}}$, both $r$ and $\frac{{\rm{d}}z}{{\rm{d}}r}$ will go to infinity, but the difference yields a finite value. To calculate $r_0$, we take the Taylor series of $r_0$ around $\tau = \tau_{\rm{max}}$. For $|E| \leq 2 \sqrt{Z_2 |F|}$, we find
\begin{equation}
	r_0 = 2 \frac{m_{\rm{cn}} \varphi }{ \theta (m_{\xi} m_{\rm{sn}}^2 - 1)} \sqrt{\xi_{\_} m_{\xi} m_+},
\end{equation}
with
\begin{equation}
	m_{\mathrm{cn}} = \cn \left(2 \varphi K(m_\eta)/\theta | m_\xi \right).
\end{equation}
For the special case $E=0$, $r_0$ is given by
\begin{equation}
	r_0 = \frac{2 \sqrt{2} (1 - \cos \beta)^{\frac{1}{4}} m_{\mathrm{cn},E=0} (\sin^2 \beta)^{\frac{1}{4}}}{\sqrt{|F|} (1 + \cos \beta)^{\frac{1}{4}} (m_{\mathrm{sn},E=0}^2 - 2)},
\end{equation}
with
\begin{equation}
	m_{\mathrm{cn},E=0} = \cn \left(
	\frac{\sqrt{\pi} (1 - \cos \beta)^{\frac{1}{4}} \Gamma(\frac{1}{4}) }
	{\phantom{\sqrt{}}2 (1 + \cos \beta)^{\frac{1}{4}} \Gamma(\frac{3}{4})} \Big|
	\frac{1}{2} \right).
\end{equation}
For $E > 2 \sqrt{Z_2 |F|}$
\begin{equation}
	r_0 = \frac{n_{\rm{cn}} \varphi }{ \psi (m_{\xi} n_{\rm{sn}}^2 - 1)} \sqrt{ \xi_{\_} m_{\xi} \frac{n_+} {1 - n_{\eta}}},
\end{equation}
with
\begin{equation}
	n_{\mathrm{cn}} = \cn \left(\varphi K(n_\eta)/\psi | m_\xi \right).
\end{equation}

With these equations, $r_0$ has been calculated as a function of the starting angle $\beta$ for $E_{\rm{exc}}$ = 5, 15, and 25\,meV in Fig. \ref{fig:vr_r0_z0_vs_beta}b. The curves show a similar structure as the curves of $v_r$ (Fig. \ref{fig:vr_r0_z0_vs_beta}a): $r_0$ increases with $\beta$, drops to zero one or multiple times and shows a complex structure near $\beta = \beta_c$. At first sight, a maximum in $r_0$ seems to coincide with $v_r = 0$ and vice versa; on closer inspection, one sees that this is only true for angles near $\beta_c$ and that the angle for which the maximum and the zero occur are not exactly the same. These small differences are caused by the influence of the Coulomb field on the electron's trajectory near the origin.

To find $z_0 = z - \frac{1}{4 a} \left( \frac{{\rm{d}}z}{{\rm{d}}r} \right)^2$, we again Taylor expand around $\tau_{\rm{max}}$. For $|E| \leq 2 \sqrt{Z_2 |F|}$, we find
\begin{equation}
	z_0 = \frac{4 \varphi^2 \left[1 + m_{\xi} (m_{\rm{sn}}^2 - 2) \right] - F \xi_{\_} m_{\xi} m_{\rm{sn}}^2}{2 F (m_{\xi} m_{\rm{sn}}^2 - 1)}.\label{eq:z0_1}
\end{equation}
For the special case $E=0$, $z_0$ is given by
\begin{equation}
	z_0 = \frac{m_{\mathrm{sn},E=0}^2 \sqrt{1 - \cos \beta}}{\sqrt{2 |F|} (m_{\mathrm{sn},E=0}^2 - 2)}.
\end{equation}
For $E > 2 \sqrt{Z_2 |F|}$, we obtain
\begin{equation}
	z_0 = \frac{4 \varphi^2 \left[1 + m_{\xi} (n_{\rm{sn}}^2 - 2) \right] - F \xi_{\_} m_{\xi} n_{\rm{sn}}^2}{2 F (m_{\xi} n_{\rm{sn}}^2 - 1)},
\end{equation}
which is equal to Eq. \eqref{eq:z0_1} with $m_{\mathrm{sn}}$ replaced by $n_{\mathrm{sn}}$.

With these equations, we have calculated $z_0$ as a function of $\beta$ for $E_{\rm{exc}}$ = 5, 15, and 25\,meV (Fig. \ref{fig:vr_r0_z0_vs_beta}c). For small excess energies, the minimum value for $z_0$ is positive; by increasing the excess energy, this value decreases linearly, is zero at $E_{\rm{exc}} = E_F$ ($E = 0$), and becomes negative for higher excess energies. For small excess energies, the electron uses a part of the energy of the acceleration field to overcome the Coulomb barrier; therefore in the limit it appears to have been accelerated over a shorter distance than it has travelled in reality, leading to a positive value of $z_0$. For $E_{\rm{exc}} = E_F$, the electron starts with just enough energy to overcome the Coulomb barrier, so for $\beta = 0$ all acceleration field energy is converted into $v_z$, so $z_0 = 0$. For higher excess energies, the electron starts with more energy then needed to overcome the barrier, leading to a larger $v_z$ than expected for $\beta = 0$, so $z_0 < 0$. The local maxima and minima in $z_0$ occur at exactly the same angles as the extreme values in $v_r$.

In the limit of $E_{\rm{exc}} \to \infty$, the effect of the Coulomb field on the electron's motion can be neglected for all $\tau$; the electron then follows a parabolic trajectory with $r_0 = (E/|F|) \sin 2 \beta$ and $z_0 = -(E/2 F) (\cos 2 \beta + 1)$ (see Appendix \ref{sec:parabol_traj}).

\section{Effective source temperature and virtual source size\label{sec:mod_src_temp}}
When calculating $v_r$ for electrons with the initial angle $\beta$ uniformly distributed over the range [0,$\beta_{\rm{c}}$), the velocity spread of the resulting electron bunch can be calculated with
\begin{align}
	\sigma_{v_{x}}^2 &= \frac{1}{2} \int_{0}^{\beta_{\rm{c}}} v_r(\beta)^2 \sin{\beta} \, {\rm{d}}\beta \big/ \int_{0}^{\beta_{\rm{c}}} \sin\beta' \, {\rm{d}}\beta' \nonumber \\
	&= \frac{1}{2 [1 - \cos{\beta_{\rm{c}}}]} \int_{0}^{\beta_{\rm{c}}} v_r(\beta)^2 \sin{\beta} \, {\rm{d}}\beta,\label{eq:sigma_vx}
\end{align}
where the term $\sin{\beta}$ creates an uniformly filled spherical shell of electrons and the factor 1/2 arises as $\sigma_{v_{x}} = \sigma_{v_{r}} / \sqrt{2}$. In Fig. \ref{fig:vr_histogram}, $\sigma_{v_{r}}$ has been indicated for the velocity distributions corresponding to $E_{\rm{exc}}$ = 5, 15, and 25\,meV. The assumption that the initial angles $\beta$ are uniformly distributed is true when the light field used for ionization is unpolarized. For polarized light fields, the model has to be extended with a polarization angle-dependent weight function; this is described in detail in Ref. \cite{Engelen_NJP_13} for the case of a linear polarization.

\begin{figure}[tp]
	\centering
	\includegraphics{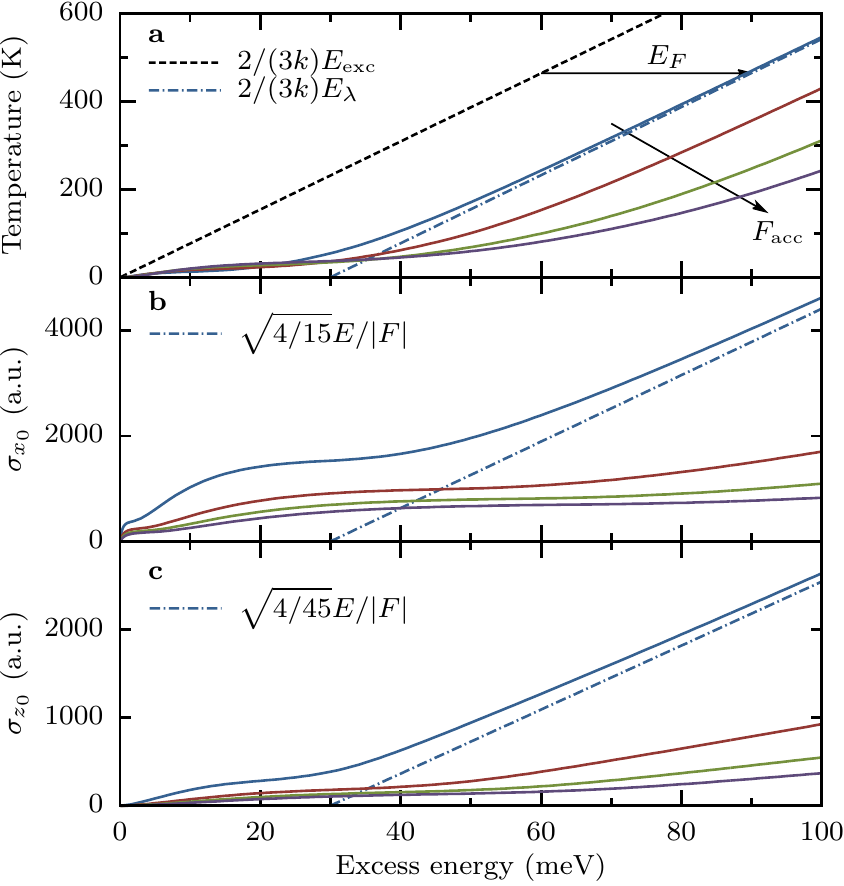}%
	\caption{Effective source temperature $T$ (a), virtual source size in the $x$-direction $\sigma_{x_0}$ (b), and effective source size in the $z$-direction $\sigma_{z_0}$ (c) as a function of excess energy $E_{\rm{exc}}$ for $F_{\rm{acc}} = -0.155, -0.370, -0.740,$ and $-1.110$\,MV/m (solid curves, respectively blue, red, green, and purple). In (a), also shown are the simple model $T = 2/(3 k) E_{\rm{exc}}$ (black dashed curve), and the simple model shifted with $E_F$ (so $T = 2/(3 k) E_{\lambda}$) for $F_{\rm{acc}} = -0.155$\,MV/m (blue dashed-dotted curve). In (b) and (c), respectively $\sigma_{x_0}$ and $\sigma_{z_0}$ are shown for $F_{\rm{acc}} = -0.155$\,MV/m when the effect of the Coulomb field on the electron's motion can be neglected (blue dashed-dotted curves)\label{fig:T_sigma_x0_sigma_z0_vs_Eexc}}
\end{figure}

With $\sigma_{v_{x}}$ and Eq. \eqref{eq:T_sigma_vx}, the effective source temperature $T$ can be calculated, which is plotted as a function of excess energy in Fig. \ref{fig:T_sigma_x0_sigma_z0_vs_Eexc}a for $F_{\rm{acc}} = -0.155, -0.370, -0.740$, and $-1.110$\,MV/m. For low excess energies, the temperature only increases slightly with increasing excess energy. Increasing the field strength does not have a significant effect on the temperature for low excess energies. For high excess energies, i.e. $E_\lambda > E_F$ or $E_{\rm{exc}} > 2 E_F$, the model curve follows the line $T = 2/(3 k_{\rm{B}}) E_\lambda$ (in SI units). This means that the temperature is fully determined by the ionization energy $E_\lambda$, thus independent of the energy shift $E_F$ due to the Stark effect. Then the electrons have so much energy that they do not feel the complex shape of the potential any more, and the transverse velocity is given by $v_r = \sqrt{2 E} \sin \beta$. At the standard simulation conditions ($F_{\rm{acc}} = -0.155$\,MV/m, $E_{\rm{exc}} = $ 5\,meV), $T = 10$\,K. This value and the trends predicted by the model agree well with measurements of the source temperature of an ultracold electron source \cite{Engelen_U_14, Taban_EPL_10}, which is based on near-threshold photoionization of laser-cooled rubidium atoms.

The effective source size in the $x$-direction $\sigma_{x_0}$ can be calculated from $r_0$ with an equation analogous to  Eq. \eqref{eq:sigma_vx}. This effective source size is equal to the virtual source size (see Appendix \ref{sec:virtual_source}). In Fig. \ref{fig:T_sigma_x0_sigma_z0_vs_Eexc}b (solid curves), $\sigma_{x_0}$ is plotted as a function of excess energy for $F_{\rm{acc}} = -0.155, -0.370, -0.740$, and $-1.110$\,MV/m. It can be seen that $\sigma_{x_0}$ decreases with increasing acceleration field strength. For the standard simulation conditions, $\sigma_{x_0} = 32$\,nm. 
In the limit of $E_{\rm{exc}} \to \infty$, $\sigma_{x_0} =\sqrt{4/15} E/|F|$, as can be derived from the expression of $r_0$ (see Appendix \ref{sec:parabol_traj}). This is illustrated for $F_{\rm{acc}} = -0.155$\,MV/m (Fig. \ref{fig:T_sigma_x0_sigma_z0_vs_Eexc}b, dashed-dotted curve).

The effective source size in the $z$-direction $\sigma_{z_0}$ can be calculated from $z_0$ with 
\begin{equation}
	\sigma_{z_0} = \sqrt{\frac{1}{1 - \cos{\beta_{\rm{c}}}} \int_{0}^{\beta_{\rm{c}}} [z_0(\beta) - \overline{z_0}]^2 \sin{\beta} \, {\rm{d}}\beta},
\end{equation}
where 
$\overline{z_0} = \int_{0}^{\beta_{\rm{c}}} z_0(\beta) \sin{\beta} \, {\rm{d}}\beta \big/ (1 - \cos{\beta_{\rm{c}}})$. In Fig. \ref{fig:T_sigma_x0_sigma_z0_vs_Eexc}c, $\sigma_{z_0}$ is plotted as a function of excess energy for $F_{\rm{acc}} = -0.155, -0.370, -0.740$, and $-1.110$\,MV/m (solid curves). For the standard simulation conditions, $\sigma_{z_0} = 4.5$\,nm. 
In the limit of $E_{\rm{exc}} \to \infty$, $\sigma_{z_0} =\sqrt{4/45} E/|F|$, which is shown for $F_{\rm{acc}} = -0.155$\,MV/m (Fig. \ref{fig:T_sigma_x0_sigma_z0_vs_Eexc}c, dashed-dotted curve).

\section{Phase space\label{sec:phase_space}}

\begin{figure*}[tp]
	\includegraphics{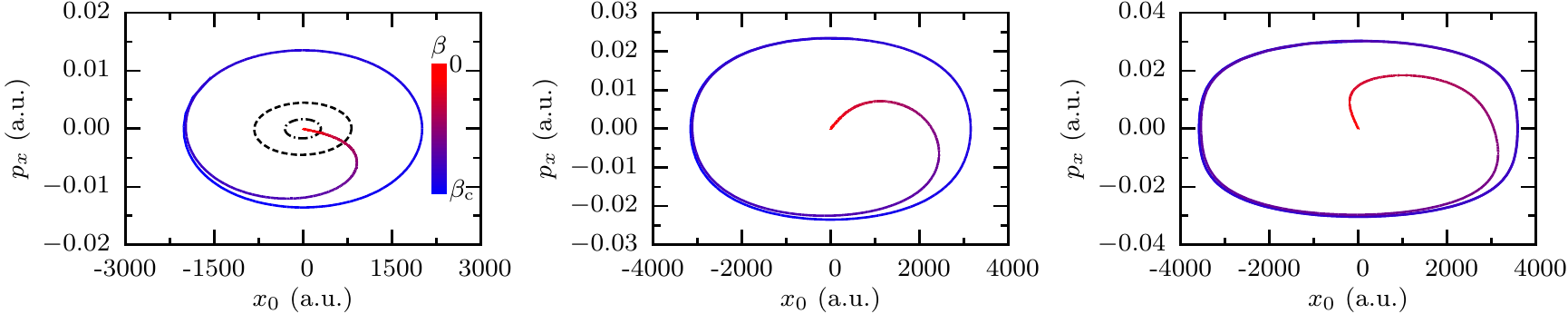}%
	\caption{Transformation of the $r_0$--$p_r$ phase space to the $x$-plane for $F_{\rm{acc}} = -0.155$\,MV/m and $E_{\rm{exc}} =$ 5\,meV (left), 15\,meV (middle), and 25\,meV (right). The colour of the curve indicates the corresponding starting angle $\beta$. In the left graph, an ellipse is plotted (dashed curve), where the product of the major and minor axis is equal to $\varepsilon_{{\rm{n}},x} / \alpha_{\rm{f}}$. The lowest possible emittance ellipse is also shown, corresponding to the Heisenberg uncertainty limit $\sigma_{x} \sigma_{p_x} = 0.5$ (dashed-dotted line).\label{fig:phase_space}}
\end{figure*}

By using the dependence of the expressions for $r_0$ and $v_r$ on $\beta$, the phase space traces can be calculated. 
The $r_0$--$p_r$ phase space distribution for $F_{\rm{acc}} = -0.155$\,MV/m and $E_{\rm{exc}} =$ 5, 15, and 25\,meV has been determined; this is shown in Fig. \ref{fig:phase_space} after transforming $r$ to the $x$-plane, as to allow axis crossings. The colour of the curve indicates the corresponding starting angle $\beta$ of the electron. For angles close to $\beta_{\rm{c}}$, the outer circle in the traces is traversed multiple times.

The volume the electron bunch occupies in phase space is a measure for the quality of the bunch. In accelerator physics, this volume is usually quantified in terms of the normalized rms emittance. In the $x$-direction, this is defined as
\begin{equation}
	\varepsilon_{{\rm{n}},x} = \alpha_{\rm{f}} \sqrt{\langle{x}^2\rangle \langle{p_x}^2\rangle - {\langle{xp_x}\rangle}^2},
\end{equation}
with $\alpha_{\rm{f}} \approx 1/137$ the fine structure constant. For $\varepsilon_{{\rm{n}},x}$ in SI units, $\alpha_{\rm{f}}$ is replaced by $1/(m c)$. The brackets $\langle .. \rangle$ indicate averaging over all electrons in the beam, with the origin of $x$ and $p_x$ chosen at the centre of the beam. For $E_{\rm{exc}} = 5$\,meV, $\varepsilon_{{\rm{n}},x} = 0.027$\,a.u.\,= 1.4\,pm rad. This emittance is illustrated in Fig. \ref{fig:phase_space} (left, dashed line), where the product of the major and minor axis of the ellipse is equal to $\varepsilon_{{\rm{n}},x} / \alpha_{\rm{f}} = 3.7$. The lowest possible emittance is also shown (dashed-dotted line), which corresponds to Heisenberg's uncertainty principle $\sigma_{x} \sigma_{p_x} = 0.5$ ($= \hbar/2$ in SI units). This shows that for the standard simulation conditions the model predicts that the electron bunch coming from this single-atom emitter is close to the Heisenberg limit. The emittance for $F_{\rm{acc}} = -0.155$\,MV/m and $E_{\rm{exc}} =$ 0.77\,meV is equal to Heisenberg limit. Obviously, in this parameter range, the classical model is no longer applicable and a full quantum-mechanical treatment is in order.

\section{Rubidium model\label{sec:mod_rubidium_model}}
In the analytical model, electrons with $\beta \ge \beta_{\rm{c}}$ cannot escape the atom, because they follow closed orbits around the atom, imposed by the $1/R$ potential. However, in experiments with electron sources based on near-threshold photoionization of laser-cooled gases, rubidium atoms are typically used \cite{Engelen_NC_13, Engelen_U_14, Engelen_NJP_13, McCulloch_NC_13, McCulloch_NP_11, Saliba_OE_12, Taban_EPL_10}. The rubidium potential is no longer a $1/R$ potential due to the inner shell electrons, so there are no closed orbits. As a result, all electrons with $E_{\rm{exc}} > 0$ will eventually leave the ion. In this section, the motion of electrons in a rubidium-Stark potential $U_{\textrm{Rb-S}}$ is studied. This potential is given by Eq. \ref{eq:pot_energy}, with $U_{\rm{ion}} = U_{\rm{Rb}}$ the rubidium potential, which is given by the sum of a model potential and the induced dipole potential \cite{Robicheaux_PRA_97}
\begin{equation}
	U_{\rm{Rb}} = -\frac{Z_L}{R} - \frac{\alpha}{2 R^4} (1 - f_3)^2.
\end{equation}
Here $Z_L = Q + f_1 + f_2 R$ is the effective charge, $Q = 1$ the ion charge, and
\begin{align}
	&f_1 = \left\{
	\begin{array}{ll}
		(Z - Q) \exp(-a_1 R) & \text{if\quad} a_1 R < 36\\
		0 & \text{if\quad} a_1 R \geq 36
	\end{array} \right. , \nonumber \\
	&f_2 = \left\{
	\begin{array}{ll}
		a_2 \exp(-a_3 R) & \qquad\;\; \text{if\quad} a_3 R < 38\\
		0 & \qquad\;\; \text{if\quad} a_3 R \geq 38
	\end{array} \right. , \nonumber \\
	&f_3 = \left\{
	\begin{array}{ll}
		\exp(-[R/R_c]^3) & \qquad \text{if\quad} R/R_c < 3\\
		0 & \qquad \text{if\quad} R/R_c \geq 3
	\end{array} \right. ,
\end{align}
with $Z$ the atomic number, $a_1$, $a_2$, and $a_3$ model function parameters, $\alpha$ the dipole polarizability of the ion core, and $R_c$ the radius of the ion core. The values that are are used in the rubidium model are shown in Table \ref{tab:Rb_model_parameters}.

\begin{table}
	\centering
  \caption{Parameters for the rubidium model (from Ref. \cite{Robicheaux_PRA_97}).}
	\begin{tabular}{ll}
	\textbf{Parameter}        & \textbf{Value} \\
	\hline
	\hline
	$Z$                       & 37 \\
	$a_1$                     & 4.28652 (5$p$ state) \\
	$a_2$                     & 9.67757 (5$p$ state) \\
	$a_3$                     & 1.74185 (5$p$ state) \\
	$\alpha$                  & 9.076 \\
	$R_c$                     & 1.0 \\
	\end{tabular}%
  \label{tab:Rb_model_parameters}%
\end{table}

When using the rubidium-Stark potential, the electron orbits can no longer be described by closed analytical expressions. The particle trajectories are calculated with the General Particle Tracer (GPT) code \cite{GPT}. Electrons are started on a spherical shell with $R = $ 5.25 $a_0$, corresponding to an electron in the 5$p$ state. They have a velocity directed radially outwards, with a kinetic energy $E_{\rm{k}} = E - U_{\textrm{Rb-S}}$. Starting with a different radius or with a velocity in a random direction has a negligible effect on the resulting electron temperature. Particles with $z < 10\,\mu$m after 1\,ns of acceleration and particles that come closer than 1\,fm to the centre of the core are removed from the simulation. From the simulated trajectories, the transverse velocity of the electrons is calculated at $z = 10\,\mu$m, where the ion potential can be neglected and the velocity has reached its asymptotic value $v_r$. When using a pure Coulomb potential in the simulation, and imposing $\beta < \beta_{\rm{c}}$, identical temperatures are obtained as with the analytical model, confirming the validity of the implementation in GPT.

\begin{figure}[tp]
	\centering
	\includegraphics{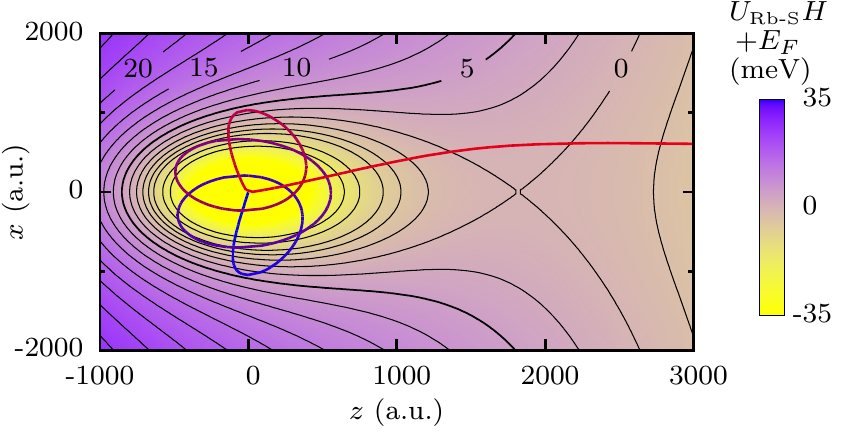}%
	\caption{Trajectory for an electron in a rubidium-Stark potential $U_{\textrm{Rb-S}}$ for $\beta = 100^{\circ}> \beta_{\rm{c}}$, calculated with the rubidium model for $E_{\rm{exc}} =$ 5\,meV and $F_{\rm{acc}} = -0.155$\,MV/m. The colour of the trajectory indicates time. After some recursions in the potential, where it is deflected when it passes close to the ion core, the electron escapes.\label{fig:potential_surface_Rb}}
\end{figure}

\begin{figure}[tp]
	\centering
	\includegraphics{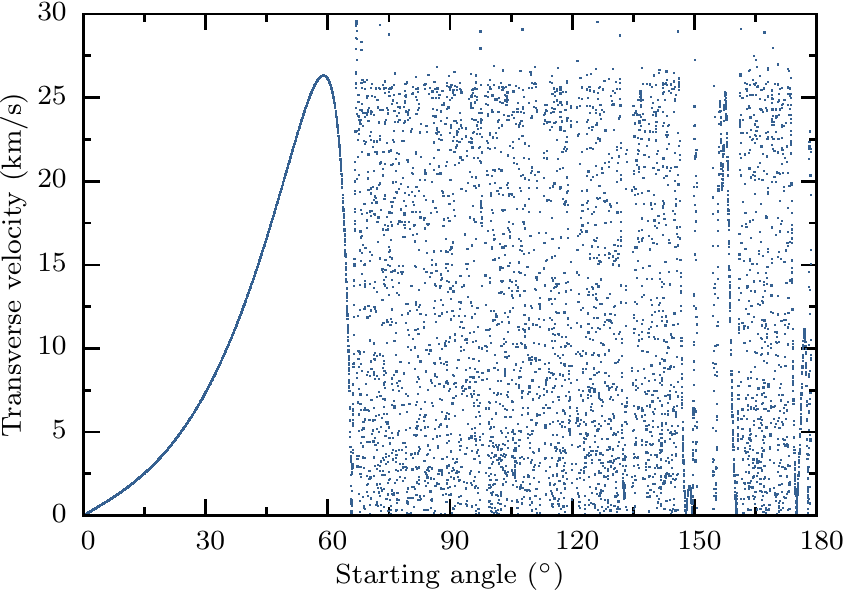}%
	\caption{Transverse electron velocity $v_r$ as a function of the electron's starting angle $\beta$, calculated with the rubidium model for $E_{\rm{exc}} =$ 5\,meV and $F_{\rm{acc}} = -0.155$\,MV/m (corresponding $\beta_{\rm{c}} = 67^{\circ}$).\label{fig:vr_Rb_vs_beta}}
\end{figure}

When this potential is used, all electrons can escape the ion. Electrons with starting angle $\beta < \beta_{\rm{c}}$ will escape the potential immediately; electrons with $\beta \geq \beta_{\rm{c}}$ will make recursions in the potential, until they scatter on the ion core. If due to this scattering event the angle between the velocity and the direction of acceleration is smaller than the critical angle, the electron will escape. An example of this is sketched in Fig. \ref{fig:potential_surface_Rb} for $\beta = 100 ^{\circ}$. Due to the deflection near the core, electrons with $\beta \ge \beta_{\rm{c}}$ will, after being able to escape the potential, follow similar trajectories as electrons with $\beta < \beta_{\rm{c}}$. To illustrate this, the transverse electron velocity is shown in Fig. \ref{fig:vr_Rb_vs_beta} as a function of starting angle. For $\beta < \beta_{\rm{c}} = 67^{\circ}$, the curve is identical to the results from the analytical model (Fig. \ref{fig:vr_r0_z0_vs_beta}a). 
Above the critical angle, $v_r$ is distributed over the same range as for $\beta < \beta_{\rm{c}}$, but there is no correlation between $v_r$ and $\beta$. This reflects the fact that the scattering from $\beta > \beta_{\rm{c}}$ to angles $\beta < \beta_{\rm{c}}$ occurs randomly. For 151$^{\circ} \leq \beta \leq 154^{\circ}$ no data points can be found. This is an artifact of the simulation conditions, in particular the choice for a cut-off radius at 1\,fm: the electrons in this range of angles move, at the start of the simulation, uphill in the potential and return to (almost) the same position as where they started from; as they then come closer than 1\,fm to the centre of the core, they are removed from the simulation.

\begin{figure}[tp]
	\centering
	\includegraphics{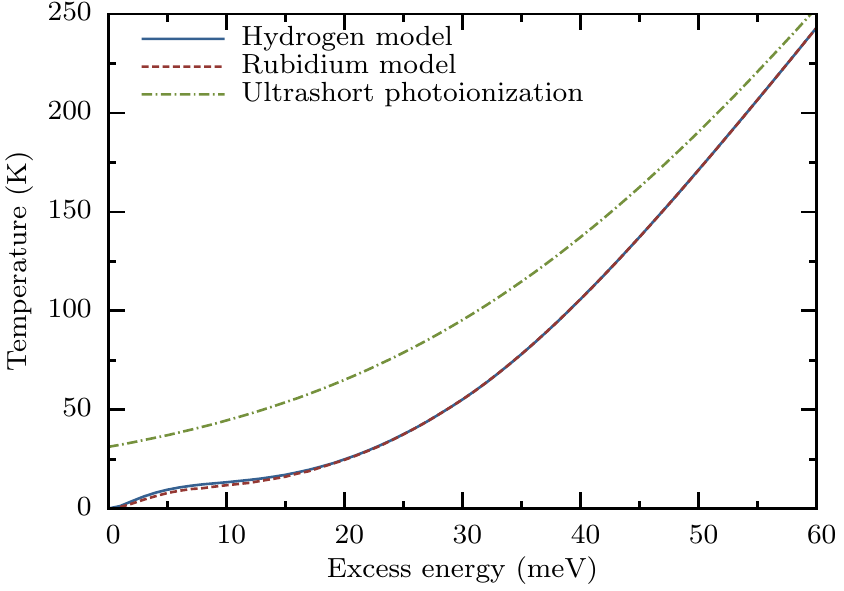}%
	\caption{Electron temperature $T$ as a function of excess energy $E_{\rm{exc}}$ for the analytical hydrogen model (blue solid curve), rubidium model (red dashed curve), and ultrashort photoionization (green dashed-dotted curve) for $F_{\rm{acc}} = -0.155$\,MV/m.\label{fig:T_vs_Eexc_data_Rb}}
\end{figure}

\begin{figure}[tp]
	\centering
	\includegraphics{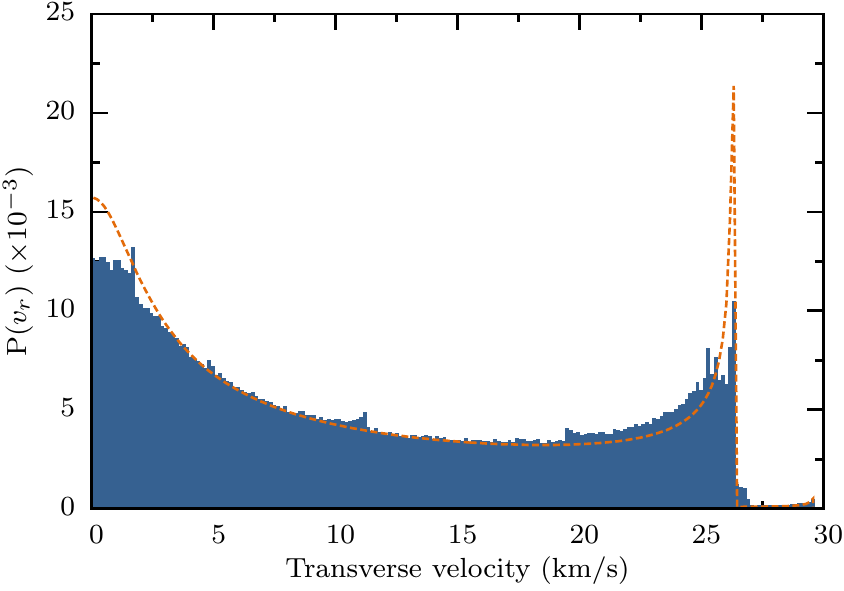}%
	\caption{Transverse velocity distribution P$(v_r)$ calculated with the rubidium model (blue solid bars) for $E_{\rm{exc}} =$ 5\,meV and $F_{\rm{acc}} = -0.155$\,MV/m. The distribution from the analytical model (orange dashed curve) is shown for comparison.\label{fig:vr_Rb_histogram}}
\end{figure}

The temperatures and virtual source sizes resulting from simulations with the rubidium-Stark potential differ only slightly from temperatures calculated with the analytical model. For the standard simulation conditions, the difference in temperature is at most 1.8\,K, see Fig. \ref{fig:T_vs_Eexc_data_Rb}. The difference arises from a slightly different distribution of the electrons over the possible trajectories, as can be seen by comparing the transverse electron velocity distribution for the analytical and rubidium model (Fig. \ref{fig:vr_Rb_histogram}). For $E_{\rm{exc}} > E_F$, all electrons in the analytical model can escape, as the critical angle $\beta_{\rm{c}} = \pi$. Therefore, the analytical model and the rubidium model yield the same temperature and source sizes for these energies.

Changing the parameters of the Rb model potential has a negligible effect on the outcome of the simulations, as long as the potential resembles a $1/r$ potential for distances larger than tens of atomic units from the ion core, as it only changes the direction of the velocity of electrons with $\beta \ge \beta_{\rm{c}}$. For example setting $f_2$ and $\alpha$ to zero, only increases the temperature for the standard simulation conditions with 0.3\%.

\section{Ultrashort photoionization\label{sec:mod_ultrashort_model}}

So far, we have assumed that in the ionization process all electrons gain the same energy, which is of course only true if the photoionization laser has a negligible spectral bandwidth. When an ultrafast laser pulse is used for ionization, the spectral bandwidth has to be taken into account in calculating the effective source temperature and virtual source size, as then electrons with a substantial spread in excess energy are created. Assuming that the number of created electrons is linear with ionization laser intensity, which has a Gaussian profile as a function of wavelength, the excess energy distribution is also Gaussian. The resulting temperature $T(\bar{E}_{\rm{exc}})$ of such a pulse can be calculated by convoluting the theoretical curve $T(E_{\rm{exc}})$ from the analytical model with a Gaussian distribution corresponding to the spectral bandwidth of the ultrashort ionization laser pulse, cut off at $E_{\rm{exc}} = 0$\,meV, as for negative excess energies no electrons are created. This leads to
\begin{equation}
	T(\bar{E}_{\rm{exc}}) = \int_0^\infty w(E_{\rm{exc}}, \bar{E}_{\rm{exc}}) T(E_{\rm{exc}}) \, {\rm{d}}E_{\rm{exc}},\label{eq:T_model_femtosecond}
\end{equation}
with $w(E_{\rm{exc}}, \bar{E}_{\rm{exc}}) =  1/C \exp \left[-(E_{\rm{exc}}-\bar{E}_{\rm{exc}})^2/2 \sigma_{E_{\rm{exc}}}^2\right]$ a normalized Gaussian weight function, $\bar{E}_{\rm{exc}}$ and $\sigma_{E_{\rm{exc}}}$ the excess energy corresponding to the central wavelength and rms wavelength spread of the ionization laser pulse respectively, and normalization constant $C = \sqrt{\pi/2} \, \sigma_{E_{\rm{exc}}}$ $\left[1 + {\rm{erf}} \left(\bar{E}_{\rm{exc}} / [\sqrt{2} \sigma_{E_{\rm{exc}}}] \right) \right]$, where erf is the error function. The source size can be calculated with an analogous equation.

In a recent experiment \cite{Engelen_NC_13}, electron bunches have been created with femtosecond laser pulses, with a wavelength around 480\,nm and an rms bandwidth $\sigma_{\lambda} = 4$\,nm, so $\sigma_{E_{\rm{exc}}} = 21.2$\,meV. The model temperature for this spectral bandwidth is plotted with a dashed-dotted curve in Fig. \ref{fig:T_vs_Eexc_data_Rb}. For $\bar{E}_{\rm{exc}} = 0$\,meV, when half of the energy in the ionization pulse can be used to create electrons, we find $T = 31$\,K. This model for the source temperature in case of ultrashort photoionization agrees well with experimental data 
\cite{Engelen_NC_13}.

\section{Conclusion\label{sec:mod_concl}}

We present an analytical model of an isolated single-atom electron source, based on a classical description of the motion of an electron in the potential of a singly charged ion and a uniform electric field used for acceleration. 

Closed analytical expressions have been derived for the asymptotic transverse velocity distribution P$(v_r)$, with which the effective source temperature $T \sim \sigma_{v_x}^2$ was determined. We find that the potential acts as a bottleneck for the electrons, and that it reduces the transverse electron velocity for small excess energies. This leads to an effective source temperature that is much lower than expected on the basis of equipartition of the available energy. For small excess energies, the temperature is largely independent of the acceleration field strength. For high excess energies ($E_{\rm{exc}} > 2 E_F$), the model curve follows the line $T = 2/(3 k) E_\lambda$. 

Furthermore, closed analytical expressions have been derived for the asymptotic parabolic electron trajectories, with which the virtual source size can be calculated. By particle tracking simulations, it was concluded that for the resulting model temperature and source size, it does not matter whether a hydrogen or a rubidium potential is used in the calculations. This shows that the model is applicable to the ultracold electron source, which is based on photoionization of laser-cooled alkali atoms. We briefly discussed what the source characteristics are when ultrashort, broadband laser pulses are used for photoionization

For a specific electric field, we have indicated the conditions required for realizing a fully coherent beam, only limited by the Heisenberg uncertainty relation. In this limit, the classical model is no longer applicable, and the system has to be treated with quantum mechanics; this is an interesting direction for further research. Furthermore, we have restricted ourselves to the description of the transverse degree of freedom in this paper. This model could be extended to the longitudinal degree of freedom, i.e. to study the longitudinal momentum spread and pulse length.

This research is supported by the Dutch Technology Foundation STW, which is part of the Netherlands Organisation for Scientific Research (NWO), and which is partly funded by the Ministry of Economic Affairs.

\appendix
\section{Electron trajectories in a uniform field\label{sec:parabol_traj}}

In a uniform electric field, the transverse electron velocity is given by $v_r = \sqrt{2 E} \sin \beta$, and the starting velocity in the longitudinal direction $v_{z,0} = \sqrt{2 E} \cos \beta$. For the electron's position, we find $r = v_r t$ and $z = v_{z,0} t - F t^2 /2$. Combining these yields $z =  v_{z,0} r / v_r - F r^2 / (2 v_r^2) = v_{z,0} r / v_r + a r^2$. We can rewrite this equation as the parabolic expression $z - z_0 = a (r - r_0)^2$, from which we find
\begin{align}
	&r_0 = (E/|F|) \sin 2 \beta, \nonumber \\
	&z_0 = -(E/2 F) (\cos 2 \beta + 1).
\end{align}

\section{Virtual source size\label{sec:virtual_source}}
The virtual source size is determined by extrapolating the asymptotic particle trajectories back to the source: where the lines cross, a virtual beam waist is formed, which usually does not coincide with the actual position of the source. The virtual source size is defined as the size of this virtual beam waist. For acceleration in a uniform electric field, as discussed in this paper, the asymptotic particle trajectories are parabolas. Consequently, the slopes of the tangents of the particle trajectories, which should be extrapolated back to the source to determine the virtual source size, depend on longitudinal position. In fact, if the vertex of the asymptotic parabolic trajectory lies at ($x = 0$, $z = 0$) and the slope of tangent is evaluated at $z = L$, then the extrapolated lines will cross exactly at ($x = 0$, $z = -L$), independent of the curvature of the parabola. By definition $z = -L$ is then the position of the virtual source. However, the relevant quantity is the size of the virtual source, not its position. The size is determined by the distribution of points where the extrapolated tangents, determined at $z = L$, of all possible asymptotic parabolic trajectories cross. If the vertex of the asymptotic parabolic trajectory lies at ($x_0$, $z_0$), then the tangents will cross at ($x = x_0$, $z = 2 z_0 - L$). The points constituting the virtual source therefore have exactly the same distribution in the $x$-direction as the distribution of vertex points, translated over a longitudinal distance $z_0 - L$. Consequently, the virtual source size is identical to the effective transverse source size. If we would define a longitudinal virtual source size, it would be twice the effective longitudinal source size.

\end{document}